\documentclass[conference]{IEEEtran}
\IEEEoverridecommandlockouts
\usepackage{amsmath,amsfonts}
\usepackage{amsthm}
\usepackage{array}
\usepackage{textcomp}
\usepackage{stfloats}
\usepackage{url}
\usepackage{microtype}
\usepackage{verbatim}
\usepackage{graphicx}
\usepackage{cite}
\usepackage{enumitem}
\usepackage{hyperref}
\usepackage{subcaption} %
\usepackage[linesnumbered,lined,algoruled,commentsnumbered]{algorithm2e}

\usepackage{caption}
\usepackage{bm}  
\usepackage{subcaption}
\usepackage[font=small,skip=0pt]{caption}
\usepackage[export]{adjustbox}
\usepackage[inkscapelatex=false]{svg}

\SetCommentSty{mycommfont}
\theoremstyle{plain}

\usepackage{units}
\newcommand{\nth}[1]{{#1}^{\text{th}}}

\newcommand{\NBS}[0]{N_{\mathrm{BS}}}

\newcommand{\RD}[0]{r_{\mathrm{RD}}}
\newcommand{\au}[0]{\theta_{\mathrm{u}}}

\newcommand{\rf}[0]{r_{\mathrm{F}}}

\usepackage{acronym}
\input{acronym}

\begin{document}
\title{Joint Motion, Angle, and Range Estimation in Near-Field under Array Calibration Imperfections
\author{Ahmed Hussain, Asmaa Abdallah, Abdulkadir Celik, and Ahmed M. Eltawil,
\\ Computer, Electrical, and Mathematical Sciences and Engineering (CEMSE) Division,
\\King Abdullah University of Science and Technology (KAUST), Thuwal, 23955-6900, KSA }
}

\maketitle

\begin{abstract}
Ultra-massive multiple-input multiple-output MIMO (UM-MIMO) leverages large antenna arrays at high frequencies, transitioning communication paradigm into the radiative near-field (NF), where spherical wavefronts enable full-vector estimation of both target location and velocity. However, location and motion parameters become inherently coupled in this regime, making their joint estimation computationally demanding. 
To overcome this, we propose a novel approach that projects the received two-dimensional space-time signal onto the angle-Doppler domain using a two-dimensional discrete Fourier transform (2D-DFT). Our analysis reveals that the resulting angular spread is centered at the target's true angle, with its width determined by the target's range. Similarly, transverse motion induces a Doppler spread centered at the true radial velocity\footnote{In this context, we refer to the angular velocity as the transverse velocity.}, with the width of Doppler spread proportional to the transverse velocity. Exploiting these spectral characteristics, we develop a low-complexity algorithm that provides coarse estimates of angle, range, and velocity, which are subsequently refined using one-dimensional multiple signal classification (MUSIC) applied independently to each parameter. The proposed method enables accurate and efficient estimation of NF target motion parameters. Simulation results demonstrate a normalized mean squared error (NMSE) of $\unit[-40]{dB}$ for location and velocity estimates compared to maximum likelihood estimation, while significantly reducing computational complexity.
\end{abstract}


\section{Introduction}
Next-generation networks are poised to utilize the abundant spectrum available in the \ac{mmWave}, and \ac{THz} frequency bands to significantly enhance spectral efficiency \cite{Jiang2021Survey}. The short wavelengths at \ac{mmWave}, and \ac{THz} facilitate deployment of thousands of antenna elements leading to \ac{UM}-\ac{MIMO} systems. As the aperture size increases, the corresponding Rayleigh distance—defined as $r_{\mathrm{RD}} = \tfrac{2D^2}{\lambda}$, where $D$ is the aperture and $\lambda$ is the wavelength—also increases. This expansion causes a significant portion of the communication to occur within the radiative \ac{NF} region. \ac{FF} propagation assumes planar wavefronts, resolving signals in the angular domain. In contrast, spherical wave propagation in the \ac{NF} enables simultaneous resolution of signal direction and distance, enabling target localization.

The estimation of target kinematic parameters in \ac{NF} sensing scenarios presents a fundamental challenge with critical practical implications \cite{hussain2025STAP}. While existing research \cite{10934779} has predominantly focused on static target localization—exploiting spherical wavefront properties to achieve range and angle estimation with minimal time-frequency resources—the moving target case remains largely unexplored despite its significance in emerging applications. 

Conventional \ac{FF} radar systems face inherent limitations in complete motion parameters estimation. Compact monostatic arrays can only detect radial velocity components due to planar wavefront approximations \cite{7801084}. This becomes particularly problematic for targets moving perpendicular to the radar's boresight (e.g., pedestrians crossing a road relative to an automotive radar), where the transverse velocity component dominates but remains completely undetectable in \ac{FF} operation. While multistatic configurations based on interferometry enable full velocity vector estimation through spatial diversity \cite{8168279}, they impose prohibitive costs in hardware complexity and synchronization precision.

The \ac{NF} regime presents a cost-effective alternative to conventional sensing approaches by exploiting spherical wavefronts that inherently encode both radial and transverse velocity components. This unique property enables even monostatic systems to estimate full two-dimensional (2D) velocity vectors without relying on complex multistatic configurations \cite{10664591,jiang2024near,meng2025near}. In this context, the work in \cite{10664591} investigates a maximum likelihood velocity estimator under the assumption of known target locations, and further extends the approach in \cite{jiang2024near} to track motion parameters across multiple \acp{CPI}. In \cite{meng2025near}, the location and velocity parameters are estimated by leveraging subarray-based variational message passing algorithm.

Despite the promise of transverse velocity estimation in \ac{NF} systems, several fundamental challenges remain. First, the intrinsic coupling between spatial and Doppler domains in \ac{NF} signal hinders the independent extraction of motion and location parameters. Second, the joint estimation of four- dimensional parameters—namely range, angle, radial velocity, and transverse velocity—results in high computational complexity. To address these challenges, we exploit the \ac{AD} response of the received \ac{NF} space-time snapshot to decouple location and velocity parameters. In \ac{NF} scenarios, a user experiences high gains across multiple \ac{DFT} bin in the angle domain, resulting in several peaks along the angle axis—a phenomenon referred to as angular spread. As shown in \cite{hussain2025near}, the true angle lies at the center of this spread, and its width is determined by the user’s range. Similarly, transverse velocity induces a spread in the Doppler frequency across multiple bins, which we refer to as Doppler support. We demonstrate that the true radial velocity lies at the center of this Doppler spread, and that its width is directly related to the target’s transverse velocity. By leveraging these relationships, we develop a low-complexity algorithm that provides coarse estimates of both location and velocity. These initial estimates are then refined using one-dimensional \ac{MUSIC} applied separately to each parameter, enabling accurate and efficient \ac{NF} motion parameter estimation. 

\begin{figure}[t!]
    \centering
    \includegraphics[width=.85\columnwidth] {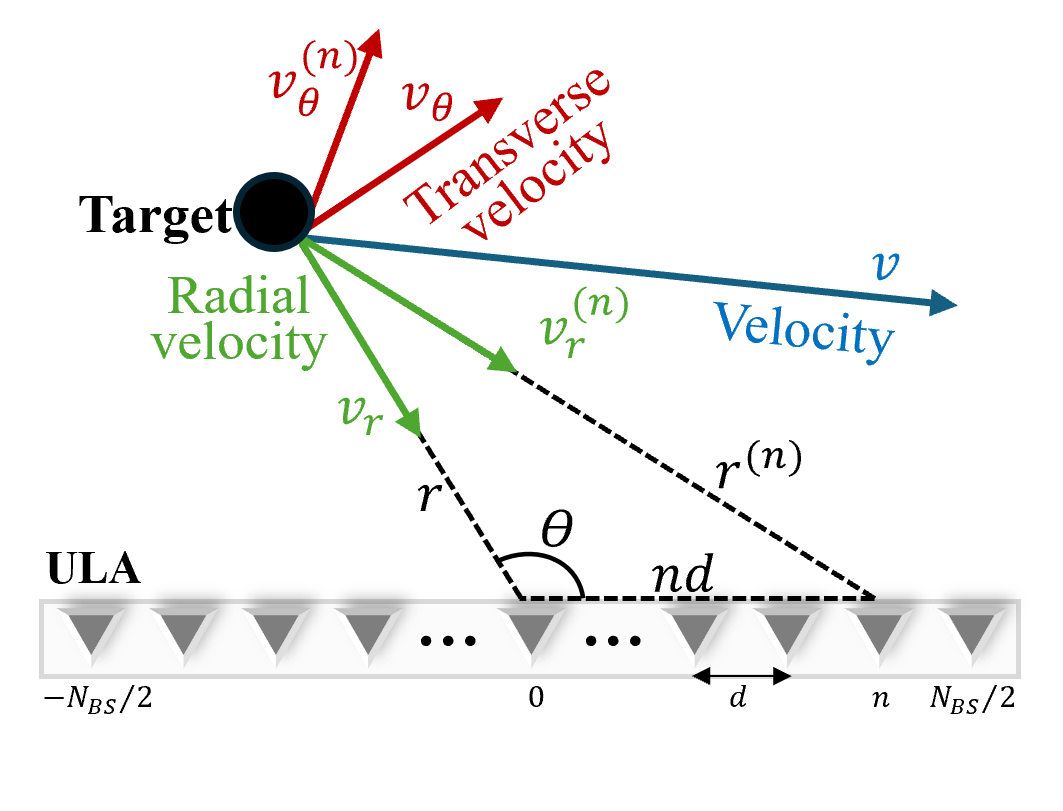}
    \setlength{\belowcaptionskip}{-18pt}
\setlength{\abovecaptionskip}{-19pt}
    \caption{\ac{NF} system model.}
    \label{system_model}
\end{figure}

\section{System Model and Problem Formulation } \label{System Model and Problem Formulation}
We consider a narrow-band system with \ac{UM}-\ac{MIMO} antenna array at the \ac{BS} and a moving point target as depicted in Fig. \ref{system_model}. The \ac{BS} is equipped with a \ac{ULA} of $\NBS$ isotropic antenna elements. For sensing operations, BS transmits $M$ symbols at a symbol rate of $f_r=\tfrac{1}{T_r}$, where $T_r$ denotes the symbol interval. The total duration of the $M$ symbols, $MT_r$ is referred to as the \ac{CPI}, during which the target parameters are assumed to remain constant.
\subsection{Space-Time Model}
\subsubsection{Spatial Steering Vector}
Let us consider a target located at a distance $r$ and angle $\theta$ from the center of the \ac{ULA}, as shown in Fig. \ref{system_model}. Based on the uniform spherical wavefront model, the \ac{NF} spatial steering vector $\mathbf{a}(\theta,r) \in \mathbb{C}^{\NBS \times 1}$ captures the phase progression across the antenna array and is given by
\begin{equation} \small
\mathbf{a} (\theta,r) = \tfrac{1}{\sqrt{\NBS}}\Big[e^{-j\nu_{c}(r^{(1)} -r)} ,\ \dots,\ e^{-j\nu_{c}(r^{(\NBS)} -r)}\Big],
\label{eqn1}
\end{equation}
\normalsize
where, $\nu = \tfrac{2\pi }{\lambda}$ is the wavenumber, and $\lambda$ is the carrier wavelength. Furthermore, $r^{(n)}$ is the distance between the target and the $n^{th}$ antenna element and computed from the law of cosines as $r^{(n)} = \sqrt{r^2 + n^2 d^2 - 2 r n d \sin{\theta}}$. We consider a fully digital architecture, where each antenna element is connected to a dedicated digital receiver via intermediate \ac{RF} hardware. This hardware may introduce phase and amplitude distortions due to various impairments, such as mismatched cable lengths, amplifier nonlinearities, and other hardware imperfections. To model these calibration errors, we introduce a diagonal matrix $\mathbf{W} \in \mathbb{C}^{\NBS \times \NBS}$, where each diagonal entry is given by $w_n = \rho_n e^{j\phi_n}$. Here, $\rho_n$ and $\phi_n$ represent the random amplitude and phase shift, respectively, associated with the $\nth{n}$ antenna element. The spatial steering vector accounting for calibration errors is expressed as $\mathbf{W}\mathbf{a}(\theta, r)$. For ease of exposition, we omit $\mathbf{W}$ in the subsequent system model derivations, but its impact will be incorporated and analyzed in the performance evaluation presented in Section \ref{Simulation Results}.

\subsubsection{Temporal Steering Vector}
We assume that the target is moving with uniform velocity $v$ with respect to the center of the \ac{ULA}, and velocity $v$ can be decomposed into radial $v_r$ and transverse components $v_{\theta}$ \footnote{$v_{\theta}$ is the angular velocity which we refer to here as transverse velocity.}. Note that $v=v^{(0)},\
v_r=v_r^{(0)},\ \text{and} \ v_\theta=v_{\theta}^{(0)}$. The local velocity at the $\nth{n}$ antenna element of the \ac{ULA} is given by $v^{(n)} = v_r^{(n)}
+ v_{\theta}^{(n)}$, where
\begin{equation}
    v_{r}^{(n)} = \tfrac{r - n d \sin \theta}{r^{(n)}} v_r, 
    \quad 
    v_{\theta}^{(n)} = \tfrac{n d \cos \theta}{r^{(n)}} v_{\theta}.
    \label{eqn2}
\end{equation}
In the \ac{FF} due to the planar wavefront, the Doppler shift is induced only due to the radial velocity component. Specifically, at large distances, $\tfrac{r - n d \sin \theta}{r^{(n)}} \rightarrow 1$ and $\tfrac{n d \cos \theta}{r^{(n)}} \rightarrow 0$. Furthermore, the Doppler shift remains the same across the entire antenna array as $v=v_{r}^{(n)} = v_{r}$. However, in the \ac{NF}, the spherical wavefront introduces variations in the Doppler shift across the array. As a result, each antenna element experiences a different Doppler shift given by $f_{t}^{(n)} = \tfrac{2v^{(n)}}{\lambda}$, where $v_n$ is the local velocity at the $n$-th element. The normalized Doppler frequency is obtained as $\omega^{(n)} = \tfrac{f_{t}^{(n)}}{f_r}$. The corresponding Doppler steering vector $\bm{b}^{(m)} \in \mathbb{C}^{\NBS \times 1}$ for the $m$-th symbol, which captures these element-wise Doppler shifts across the $\NBS$ antennas, is given by
\begin{equation}\small
    \bm{b}^{(m)}(v_r, v_{\theta})\!\! =\!\! 
    \begin{bmatrix}
        e^{-j \pi m\omega^{(1)}}, \ e^{-j \pi m\omega^{(2)}}, \ \cdots, \ e^{-j \pi m\omega^{(\NBS)}}
    \end{bmatrix}.
    \label{eqn3}
\end{equation}\normalsize
To account for the Doppler shift across all $M$ symbols, we concatenate $\bm{b}^{(m)}, \forall m \in [1, M],$ to obtain the Doppler steering matrix $\bm{\mathcal{B}} \in \mathbb{C}^{ { \NBS \times M}}$ as
\begin{equation}\small
    \bm{\mathcal{B}}(v_r, v_{\theta})\!\! =\!\!
    \begin{bmatrix}
        \bm{b}^{(1)}(v_r, v_{\theta}) & \bm{b}^{(2)}(v_r, v_{\theta}) & \dots & \bm{b}^{(M)}(v_r, v_{\theta})
    \end{bmatrix}.
    \label{eqn4}
\end{equation}
\normalsize
\subsubsection{Target Steering vector }
To enable joint processing of signals across both spatial and temporal dimensions, the \textit{space-time steering matrix} $\bm{\mathcal{V}} \in \mathbb{C}^{ { \NBS \times M}}$ is constructed as
\small
\begin{equation}
    \bm{\mathcal{V}}(\theta,r,v_r, v_{\theta}) = \sqrt{\xi_t}\left(\bm{\mathcal{A}}(\theta, r) \odot \bm{\mathcal{B}}(v_r, v_{\theta})\right),
    \label{eqn5}
\end{equation}
\normalsize
where $\bm{\mathcal{A}} = \bm{a} \otimes \bm{1}_{1 \times M}$ and $\odot$ denotes the Hadamard (element-wise) product. Moreover, the target \ac{SNR} is given by 
\small
\begin{equation}
\xi_t = \tfrac{P_T G^2 \lambda^2 \sigma_{RCS}}{(4\pi)^3 r^4 },
\label{eqn6}
\end{equation}
\normalsize
where, $P_t$ denotes transmit power; $G$ represents the antenna gain; and $\sigma_{\mathrm{RCS}}$ represents the target \ac{RCS}. 

\subsection{Problem Formulation}
We consider space-time snapshot for a single moving target such that its range remains fixed during one \ac{CPI} and is moving with uniform velocity. Let $\mathbf{s}(m)=\left[s_{1}(m), \ \ldots, \ s_{\NBS}(m)\right] \in \mathbb{C}^{\NBS \times 1}$ denote the transmit signal of the BS at time index $m$, with $s_{n}(m)$ representing the transmit signal of the $\nth{n}$ antenna. Then, the received echo reflected from the single target during one \ac{CPI} is given by
\begin{equation}\small
\mathbf{Y}=\mathbf{X}(\theta,r,v_r, v_{\theta})+\mathbf{Z}, 
\label{eqn7}
\end{equation}
\normalsize
where, $\mathbf{X}(\theta,r,v_r, v_{\theta})=\left[\bm{\mathcal{V}}_{1} \mathbf{s}(1), \ \ldots, \ \bm{\mathcal{V}}_{M} \mathbf{s}(M)\right]$, $ \mathbf{Z}=$ $[\mathbf{z}(1), \ \mathbf{z}(2), \ \ldots, \ \mathbf{z}(M)]$, and $z_{n}(m) \sim \mathcal{C N}\left(0, \sigma^{2}\right)$ denotes the complex Gaussian noise. The unknown location and velocity parameters can be estimated by maximizing the likelihood, which can formulated as the following optimization problem
\begin{equation}\small
(\hat{\bm{\psi}}, \hat{\boldsymbol{v}})=\arg \min _{\bm{\psi}, \boldsymbol{v}}\|\mathbf{Y}- \mathbf{X}(\bm{\psi}, \boldsymbol{v})\|_{F}^{2},
\label{eqn8}
\end{equation}
\normalsize
where $\hat{\bm{\psi}} = [\hat{\theta}, \hat{r} ]$ and $\hat{\boldsymbol{v}}=\left[\hat{v}_{r}, \hat{v}_{\theta}\right]$ denote the parameters to be estimated. The Doppler phase term in \eqref{eqn3} is inherently coupled with the antenna dimension as shown in \eqref{eqn2}, which necessitates a joint 4{D} parameter estimation over the variables $[\theta, r, v_r, v_{\theta}]$. Therefore, solving the optimization problem in \eqref{eqn8} is intractable due to exponential complexity with grid size. To address this, we propose a low-complexity 2D-\ac{DFT}-based method that reduces the original 4D search to an efficient 1D linear search.
\section{Angle-Doppler Response of a NF Target}
In this section, we analyze the \ac{AD} response of a \ac{NF} target to extract its motion parameters. We begin by examining the angular spectrum of a \ac{NF} target, followed by an analysis of its Doppler spectrum using a \ac{DFT}-based approach. 
The \ac{DFT} framework facilitates the decoupling of spatial and temporal dimensions, enabling parallel and scalable parameter estimation.
\vspace{-2 mm}
\subsection{Angular Spectrum}
The \ac{AoA} of a \ac{LoS} \ac{FF} user can be accurately estimated by applying the \ac{DFT} to the received spatial steering vector. However, a \ac{NF} target exhibits angular spreading in the \ac{DFT} spectrum, which depends on its distance from the \ac{BS}. This discrepancy arises from the difference in wavefront curvature: \ac{FF} signals are characterized by planar wavefronts, while \ac{NF} signals exhibit spherical wavefronts. These spherical wavefronts can be interpreted as a superposition of multiple planar components. To illustrate this further, consider a BS serving a \ac{LoS} \ac{NF} user located at angle $\theta_u$ and distance $\rf$. The element-wise propagation distance $\rf^{(n)}$ can be approximated using a second-order Taylor expansion as
\begin{equation}\small
\rf^{(n)} \approx \rf + nd\sin(\theta)- \tfrac{1}{2\rf}n^2d^2\cos^2(\theta).
\label{eqn9}
\end{equation}
\normalsize
Given a \ac{DFT} codebook $\mathbf{f}$, the angle domain response of a \ac{NF} target located at angle $\au$ and range $\rf$ can be evaluated as
\begin{equation} \small
\begin{aligned}
&\mathcal{G}_{\mathrm{ang}}(\au,\rf;\theta_n) ={\left| { \mathbf{a} ^{\mathsf{H}} (\au,\rf) \mathbf{f} (\theta_n)} \right|}^2,\\
&\approx
\tfrac{1}{\NBS^2}\left|\sum_{-\NBS/2}^{\NBS} e^{-j\pi \{ n^2(\tfrac{d\cos^{2}\au}{2\rf}) - n(\sin\au - \sin\theta_n)\} } \right|^2,
\end{aligned}
\label{eqn10}
\end{equation}
\normalsize
where, $\mathcal{G}_{\mathrm{ang}}(\au,\rf;\theta_n)$ is the absolute normalized \ac{DFT} gain at angle $\theta_n$. The above expression can be further simplified in the form of Fresnel functions as \cite{hussain2025near} 
\begin{equation}\small
\begin{aligned}
\mathcal{G}_{\mathrm{ang}}(\au,\rf;\theta_n) 
&\approx
\left|\tfrac{\overline{\mathcal{C}}\left(\gamma_{1}, \gamma_{2}\right)+j \overline{\mathcal{S}}\left(\gamma_{1}, \gamma_{2}\right)}{2 \gamma_{2}}\right|^2, 
\end{aligned}
\label{eqn11}
\end{equation}
\normalsize

$\overline{\mathcal{C}}\left(\gamma_{1}, \gamma_{2}\right) \equiv \mathcal{C}\left(\gamma_{1}+\gamma_{2}\right)-\mathcal{C}\left(\gamma_{1}-\gamma_{2}\right)$ and $\overline{\mathcal{S}}\left(\gamma_{1}, \gamma_{2}\right) \equiv$
 $\mathcal{S}\left(\gamma_{1}+\gamma_{2}\right)-\mathcal{S}\left(\gamma_{1}-\gamma_{2}\right)$, where $\gamma_{1}=\sqrt{\tfrac{\rf}{d\cos^{2}\au}}(\sin\theta_n-\sin\au) $ and $\gamma_{2}=\tfrac{\NBS}{2} \sqrt{\tfrac{d\cos^{2}\au}{\rf}}$.

For a given target, the parameter $\gamma_2$ remains constant, whereas the gain function $\mathcal{G}(\au,\rf;\theta_n)$ primarily varies with $\gamma_1$, which is dependent on the observation angle $\theta_n$. \textbf{As a result, the gain pattern of a NF target exhibits angular spread centered around the user's true angle $\au$}.  
We define the angular spread $\Omega_\mathrm{3dB}$ as the set of angles where the \ac{DFT} gain remains within 3 dB of the peak value, expressed mathematically as
\begin{equation}\small
\begin{aligned}
\Omega_\mathrm{3dB} \stackrel{\Delta}{=} \left\{ \theta_n \mid \mathcal{G}(\au,\rf;\theta_n) > 0.5 \max_{\theta_n} \mathcal{G}(\au,\rf;\theta_n) \right\}.
\end{aligned}
\label{eqn12}
\end{equation}
\normalsize
\textbf{The width of $\Omega_\mathrm{3dB}$ is uniquely linked to the user's range $\rf$ — it is broader for users located closer to the array and narrows down as the range increases.} In particular, noticeable angular spreading occurs only when the user is located within the \ac{EBRD}, characterized by $\rf < \tfrac{\RD \cos^2\theta}{10}$ for a \ac{ULA} \cite{hussain2025near}. This boundary not only captures the onset of angular spread but also delineates the region where key \ac{NF} effects manifest, such as finite beam depth and increased spatial degrees of freedom. Beyond the \ac{EBRD}, these \ac{NF} phenomena cease to exist, and the the system behavior is equivalent to \ac{FF}.
\vspace{-2 mm}
\subsection{Doppler Spectrum}
In this subsection, we analyze how the Doppler frequency of a \ac{NF} target varies across the spatial dimension. Building on the angular spectrum analysis in the previous subsection, our objective is to characterize the Doppler spectral features and leverage them to estimate the target radial and transverse velocity components.

The local velocity expressions in \eqref{eqn2} are derived from the exact spherical propagation model but result in complicated analytical expressions. To simplify the analysis, we employ a second-order Taylor approximation of the range profile $\rf^{(n)}$ from \eqref{eqn9}, which yields the approximate expressions for the local radial and transverse velocities as
$
v_r^{(n)} \approx v_r \left(1 + \tfrac{(n d)^2 \cos^2 \theta_u}{2 \rf^2} \right), \quad
v_\theta^{(n)} \approx v_\theta \left( \tfrac{n d  \cos \theta_u}{\rf} + \tfrac{(n d)^2 \sin \theta_u \cos \theta_u}{\rf^2} \right).
$
By neglecting the higher-order $\mathcal{O}(1/\rf^2)$ terms, we further simplify the local velocity expressions as $v_r^{(n)} \approx v_r, \quad
v_{\theta}^{(n)} \approx \tfrac{n d  \cos \theta_u}{\rf} v_{\theta}.$ Based on this approximation, we can express the Doppler-domain gain at angle $\theta_n$ by evaluating the inner product between the \ac{NF} Doppler steering vector $\mathbf{b}(v_r, v_\theta)$ and a \ac{FF} \ac{DFT} codebook $\mathbf{f}$ as
\begin{equation} \small
\begin{aligned}
&\mathcal{G}_{\mathrm{dopp}}(v_r,v_{\theta};\theta_n) = \left| \mathbf{b}^{\mathsf{H}}(v_r,v_{\theta}) \, \mathbf{f}(\theta_n) \right|^2, \\
&\approx \tfrac{1}{\NBS^2} \left| \sum_{n=-\NBS/2}^{\NBS/2} 
e^{-j\pi \left[ -n\left( \tfrac{v_{\theta} \cos\theta_u}{\rf f_r} - \sin\theta_n \right) + v_r \right]} \right|^2.
\end{aligned}
\label{eqn13}
\end{equation}
\normalsize
In the above expression, the phase shift varies linearly across the antenna array, where the phase slope with respect to the antenna index $n$ is governed by the term $\tfrac{v_{\theta} \cos\theta_u}{\rf f_r} - \sin\theta_n$. This slope is directly proportional to the transverse velocity $v_{\theta}$, scaled by $\cos\theta_u$, and inversely proportional to the user range $\rf$. Analogous to the angular spread defined in \eqref{eqn12}, we define the Doppler spread $\omega_\mathrm{3dB}$ as the set of Doppler bins where the gain remains within 3 dB of its maximum value. \textbf{This Doppler spread (or phase slope) is thus uniquely determined by the transverse velocity $v_{\theta}$.} Moreover, the entire gain pattern is shifted due to the radial velocity component $v_r$. \textbf{Consequently, the center of the Doppler gain pattern—i.e., the midpoint of the Doppler spread—directly corresponds to the radial velocity $v_r$.}

\begin{figure}[t!]
    \centering
    \includegraphics[width=\columnwidth] {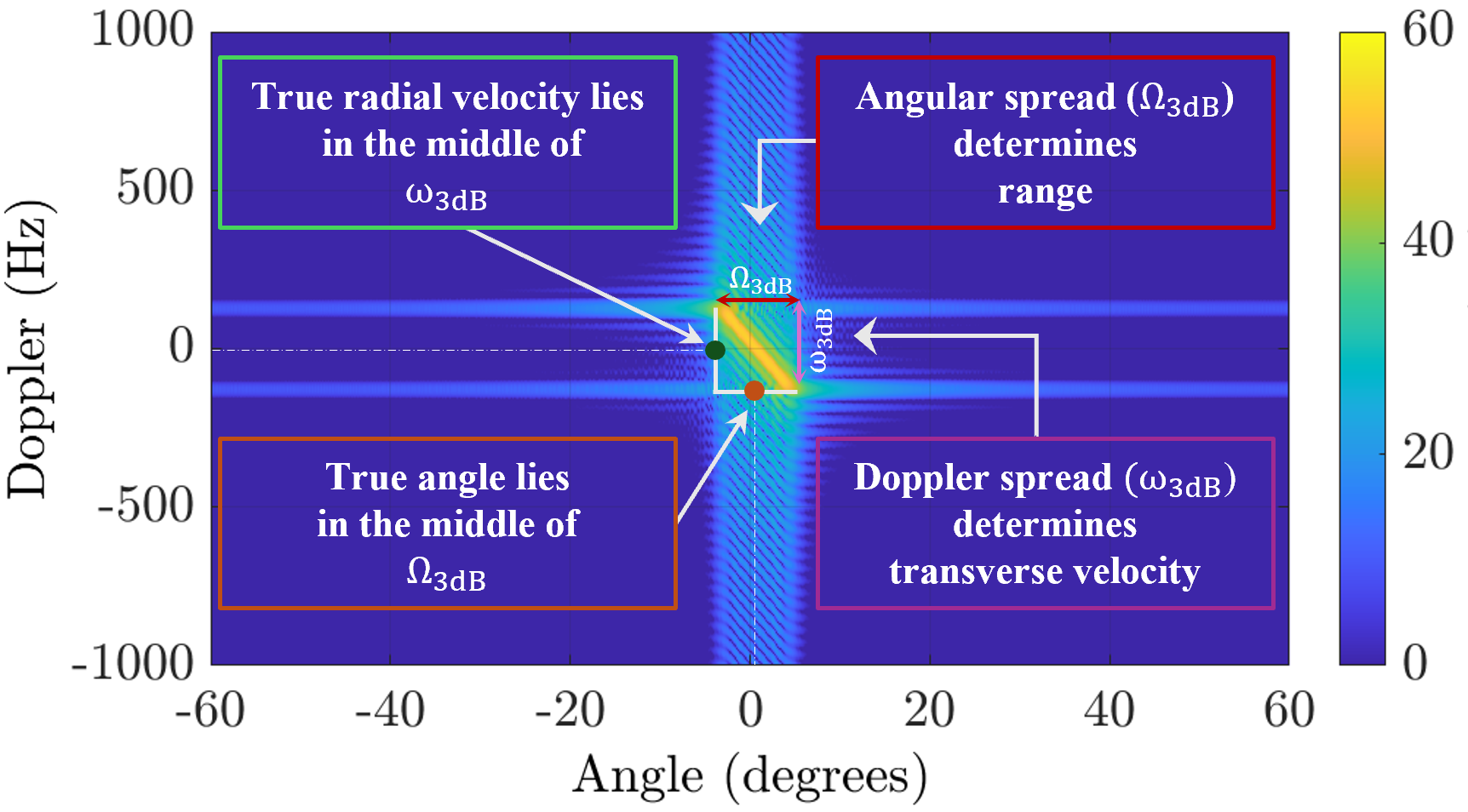}
    \setlength{\belowcaptionskip}{-15pt}
\setlength{\abovecaptionskip}{-15pt}
    \caption{Angle Doppler response for a NF target with parameters as $[\theta=0, \ \rf =\RD/50 \ \text{m}, \ v_r = 0, \quad  v_{\theta}=10 \ \text{m/s}$].}
    \label{fig:angle_Doppler}
\end{figure}

To explain the preceding analysis visually, Fig. \ref{fig:angle_Doppler} illustrates the \ac{AD} response of a \ac{NF} target characterized by the parameters $[\theta = 0^{\circ},\ \rf = \RD/50\ \text{m},\ v_r = 0,\ v_{\theta} = 10\ \text{m/s}]$. As observed, the angular spread is centered around the true angle $\theta = 0^{\circ}$, while the Doppler spread is centered around the true radial velocity $v_r = 0$. Notably, the extent of the angular and Doppler spreads is indicative of the target’s range and transverse velocity, respectively. These characteristics can be exploited through correlative interferometry techniques, which are elaborated upon in the subsequent section.
\vspace{-2 mm}
\section{Parameters Estimation}
In this section, we outline a two-stage approach for estimating the target motion parameters. In the first stage, coarse estimates are obtained using the \ac{AD} analysis introduced earlier. These initial estimates are then refined in the second stage by applying 1{D} \ac{MUSIC}.

\vspace{-1 mm}
\subsection{DFT-based Parameter Estimation}
In the first step, 2{D}-\ac{DFT} of the received signal is computed to obtain coarse estimates of the location and motion parameters, as detailed in the following paragraph:

\subsubsection{Angle and Radial Velocity}
The coarse angle $\tilde{\theta}$ of the target is estimated by computing the median of the angular spread as
\small
\begin{equation}
\tilde{\theta} \approx \operatorname{Med}\left(\hat{\Omega}_\mathrm{3dB}\right).
\label{eqn14}
\end{equation}
\normalsize
Likewise, the coarse radial velocity $\tilde{v}_r$ is estimated by computing the median of the Doppler spread as
\small
\begin{equation}
\tilde{v}_r \approx \operatorname{Med}\left(\hat{\omega}_\mathrm{3dB}\right).
\label{eqn15}
\end{equation}
\normalsize

\subsubsection{Range and Transverse Velocity}
Building upon the previous analysis, where the angular and Doppler spreads were shown to encode information about the target’s range and transverse velocity, respectively, we propose a correlative interferometry-based approach for estimating these parameters. Specifically, we construct two lookup tables: ${\mathcal{K}_{a}}(\theta_i, r_j)$, which captures the relationship between angular spread and range across different angles, and ${\mathcal{K}_{v}}(v_{r,i}, v_{\theta,j})$, which maps Doppler spread to radial and transverse velocity pairs.
These tables are populated offline by simulating targets over a predefined grid of positions and velocities, computing the corresponding 2D-\ac{DFT} responses, and extracting the $3$-dB angular and Doppler spreads, denoted by $\Omega_{\mathrm{3dB}}$ and $\omega_{\mathrm{3dB}}$, for each scenario. The resulting values are stored in the respective lookup tables, which only need to be computed once and can be reused during real-time estimation.
During operation, the \ac{BS} measures the angular spread $\hat{\Omega}_{\mathrm{3dB}}$ and Doppler spread $\hat{\omega}_{\mathrm{3dB}}$ from the noisy received signal. These measurements are then correlated with the entries in the lookup tables to estimate the target’s range $\tilde{r}$ and transverse velocity $\tilde{v}_{\theta}$ as 
\begin{equation}\small
\begin{aligned}    
  \tilde{r} &= \arg\max_{r_j} \left[ \cos\left( \hat{\Omega}_{\mathrm{3dB}} - \mathcal{K}_{a}(\theta_i, r_j) \right) \right], \\
  \tilde{v}_{\theta} &= \arg\max_{v_{\theta,j}} \left[ \cos\left( \hat{\omega}_{\mathrm{3dB}} - \mathcal{K}_{v}(v_{r,i}, v_{\theta,j}) \right) \right].
\end{aligned}
\label{eqn16}
\end{equation}
\normalsize
\subsection{MUSIC-based High-Resolution Parameter Estimation}
To achieve high-accuracy estimation, the super-resolution \ac{MUSIC} algorithm is employed to refine each location and velocity parameter separately. The refinement is conducted sequentially: the angle is first refined, followed by the range. The resulting location estimates are subsequently utilized to refine the velocity parameters, beginning with the radial component and then the transverse component.

At each stage, the signal subspace is extracted from the sample covariance matrix $\mathbf{R} = \mathbf{Y}\mathbf{Y}^H$, where $\mathbf{Y}$ is the received data matrix. The noise subspace $\mathbf{U}_n$ is obtained via eigenvalue decomposition. The parameter-specific steering vector $\mathbf{a}(\cdot)$ is then scanned over a fine grid, and the estimate is obtained by maximizing the \ac{MUSIC} spectrum 
$\hat{x} = \arg\max_{x} \left( \tfrac{1}{\|\mathbf{U}_n^H \mathbf{a}(x)\|^2} \right)$.

\begin{figure*}[!t]
  \centering
  \begin{minipage}[b]{0.24\textwidth}
    \centering
    \includegraphics[width=\textwidth]{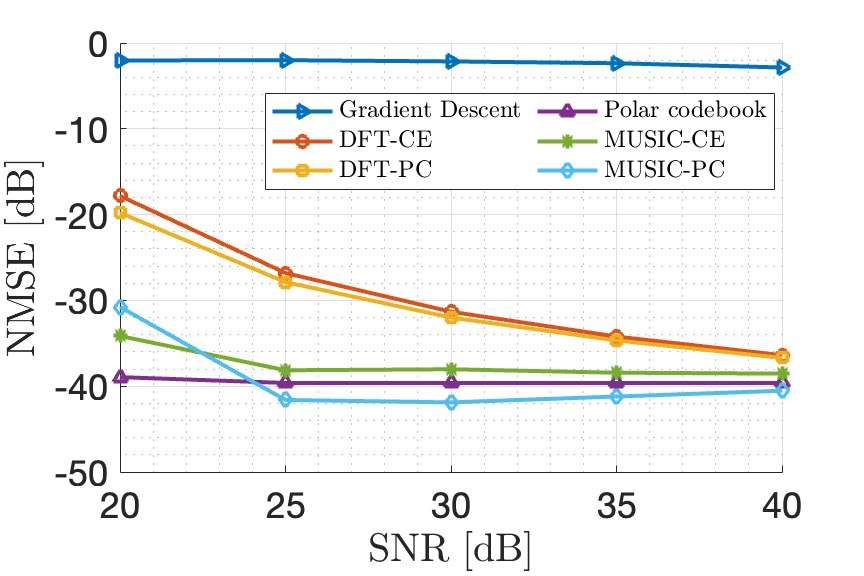}
    \caption{NMSE of estimated angle vs. \ac{SNR}.}
    \label{fig3}
  \end{minipage}%
  \hspace{0.5mm}
  \begin{minipage}[b]{0.24\textwidth}
    \centering
    \includegraphics[width=\textwidth]{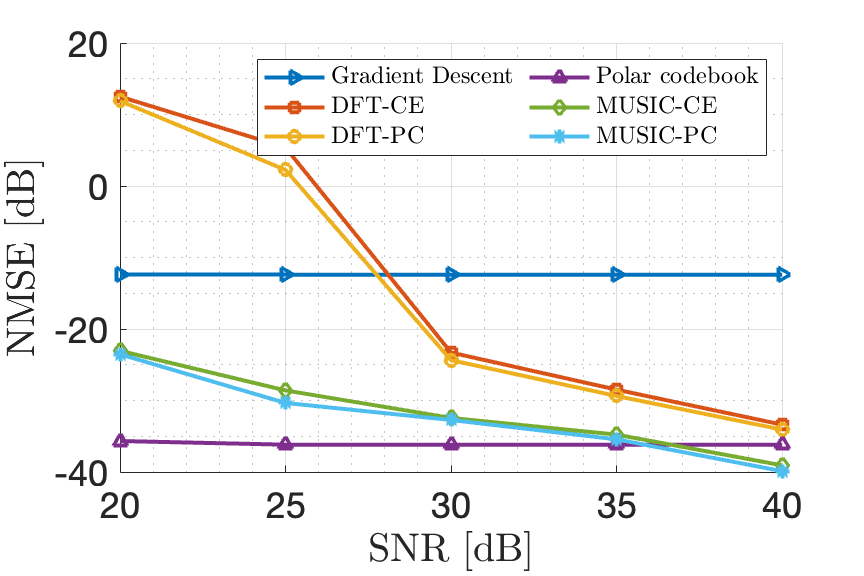}
    \caption{ NMSE of estimated range vs. \ac{SNR}.}
    \label{fig4}
  \end{minipage}
  \hspace{.5mm}
  \begin{minipage}[b]{0.24\textwidth}
    \centering
    \includegraphics[width=\textwidth]{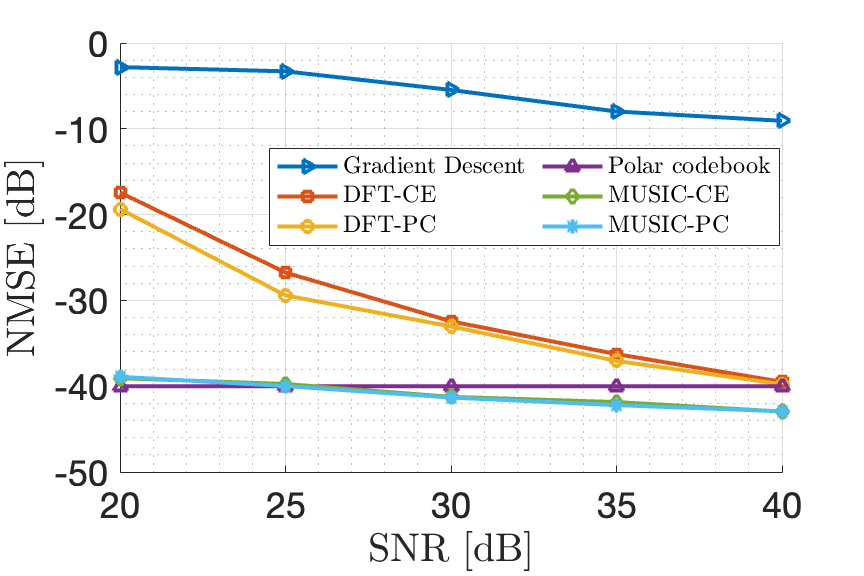}
    \caption{NMSE of estimated radial velocity vs. \ac{SNR}.}
    \label{fig5}
  \end{minipage}
  \begin{minipage}[b]{0.24\textwidth}
    \centering
    \includegraphics[width=\textwidth]{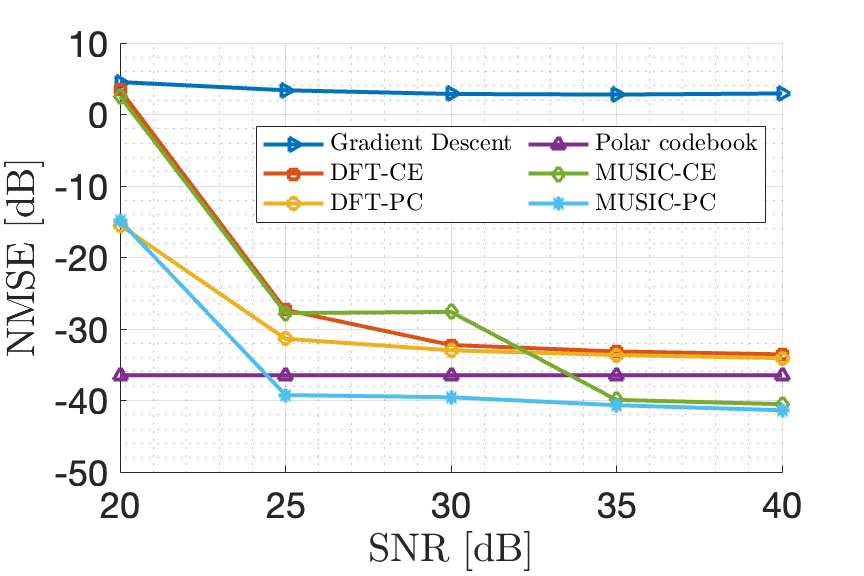}
    \caption{NMSE of estimated transverse velocity vs. \ac{SNR}.}
    \label{fig6}
  \end{minipage}
  \vspace{-0.2in}
\end{figure*}

\section{Simulation Results} \label{Simulation Results}
In this section, we evaluate the performance of the proposed parameter estimation algorithm through numerical simulations. The \ac{BS} is equipped with $N_{\text{BS}} = 256$ antennas and operates at a carrier frequency of $28~\text{GHz}$. For this configuration, the Rayleigh distance is $\RD = \unit[350]{m}$, and the \ac{EBRD} is given by $35\cos^2\theta$. The \ac{BS} transmits $M = 32$ symbols at a rate of $f_r = 5~\text{kHz}$. Array calibration errors are modeled by introducing random phase offsets uniformly distributed within $\mathcal{U}[0, \pi/36]$ and amplitude errors in dB drawn from $\mathcal{U}[0, 1]$. The true target parameters are set as follows: range and angle are $r = \RD / 50$ and $\theta = \pi/12$, respectively; the radial and transverse velocities are $v_r = 10~\text{m/s}$ and $v_\theta = 8~\text{m/s}$.

We compute the \ac{NMSE} to evaluate the estimation accuracy for range, angle, radial velocity, and transverse velocity. The \ac{NMSE} for each unknown parameter $ x \in \{ r, \theta, v_{r}, v_{\theta} \} $ is evaluated using the expression 
$ \text{NMSE}_x = \tfrac{\mathbb{E}\left[ |x - \hat{x}|^2 \right]}{\mathbb{E}\left[ |x|^2 \right]} $, 
where \( \hat{x} \) denotes the estimate of the true value \( x \). We plot \ac{NMSE} against \ac{SNR} of the processed signal, where an additional gain of $10 \log_{10}(M \times \NBS) = 39 \ \text{dB}$ is achieved due to 2D DFT processing. All the presented results are based on $1000$ iterations. We refer to the {\ac{DFT}}-based estimates under perfect calibration as \ac{DFT}-PC, while \ac{DFT}-CE denotes \ac{DFT}-based estimates in the presence of calibration errors. Similarly, \ac{MUSIC}-PC and \ac{MUSIC}-CE represent the refined estimates obtained using the \ac{MUSIC} algorithm under perfect calibration and with calibration errors, respectively. We compare our approach with the following benchmark schemes:
\begin{itemize} 
\item \textbf{Maximum Likelihood Estimation}: We adopt the gradient-based method presented in \cite{10664591} to obtain the maximum likelihood estimates of the target motion parameters.

\item \textbf{Polar Codebook}: This scheme utilizes separate codebooks for location and velocity estimation. For range and angle estimation, the polar codebook from \cite{ahmed2024near} is employed. For velocity estimation, a distinct 2D codebook is constructed over a grid of radial and transverse velocities, assuming that the angle and range parameters are known. 
\end{itemize}

Fig.~\ref{fig3} and Fig.~\ref{fig4} show the NMSE performance of angle and range estimation for the proposed method compared to benchmark schemes. The over-sampled polar codebook, constructed using $5000$ grid points, serves as an upper bound on estimation accuracy. The proposed \ac{MUSIC}-PC and \ac{MUSIC}-CE algorithms refine the coarse estimates provided by \ac{DFT}-PC and \ac{DFT}-CE, respectively, and closely approach the polar codebook performance at high \acp{SNR}. In contrast, the gradient descent method exhibits the highest estimation error, primarily due to joint optimization over coupled position and velocity parameters, which results in complex local optima in a high-dimensional search space. 

Fig.~\ref{fig5} and Fig.~\ref{fig6} further illustrate the NMSE performance for radial and transverse velocity estimation, respectively. Once again, \ac{MUSIC}-PC and \ac{MUSIC}-CE demonstrate superior accuracy, closely tracking the polar codebook solution. Specifically, the proposed \ac{MUSIC}-PC and \ac{MUSIC}-CE methods achieve an NMSE of $\unit[-40]{dB}$ across angle, range, radial, and transverse velocity estimations at high SNR levels.

It is important to note, however, that the initial \ac{DFT}-based estimates for both range and transverse velocity degrade as the user's distance from the BS increases, which may impose a heavier computational load on the \ac{MUSIC} algorithm during the refinement stage. Moreover, the proposed estimation framework is valid only within the \ac{NF} region bounded by the \ac{EBRD}, beyond which angular and Doppler spreads diminish, rendering the method inapplicable.

The computational complexities of the gradient-based method, polar codebook, and the proposed \ac{DFT}+\ac{MUSIC} algorithm are $\mathcal{O}(TG^4 \NBS M)$, $\mathcal{O}(2G^2 \NBS M)$, and $\mathcal{O}(\NBS M (\log \NBS + \log M))$ + $\mathcal{O}((\NBS M)^3 + G \NBS M)$, respectively. Here, $G$ represents the grid points per dimension and $T$ denotes the number of iterations. The proposed algorithm is more scalable as it avoids the $G^4$ complexity. The $(\NBS M)^3$ term arises from the singular value decomposition (SVD) computation in \ac{MUSIC}. 

\section{Conclusion}
In this work, we have proposed a two-stage approach for efficient motion parameter estimation in \ac{NF} \ac{UM}-\ac{MIMO} systems. In the first stage, the received signal was projected onto the \ac{AD} domain using a 2D-DFT to obtain coarse estimates of angle, range, and both radial and transverse velocities. In the second stage, these estimates were refined using one-dimensional \ac{MUSIC} applied separately to each parameter. Future work may focus on reducing the complexity of the refinement stage by exploring alternative subspace methods and developing a single-stage deep learning solution that leverages the spectral characteristics of the \ac{AD} response of the \ac{NF} target.

\bibliographystyle{IEEEtran}
\bibliography{IEEEabrv,my2bib}
\end{document}